\documentclass[twocolumn,showpacs,superscriptaddress,amsmath,amssymb]{revtex4}

\usepackage{float,epsfig}

\begin{document}

\title{Shell-model structure of exotic $^{135}$Sb}
\author{L. Coraggio} 
\author{A. Covello} 
\author{A. Gargano} 
\author{N. Itaco}
\affiliation{Dipartimento di Scienze Fisiche, Universit\`a
di Napoli Federico II, \\ and Istituto Nazionale di Fisica Nucleare, \\
Complesso Universitario di Monte  S. Angelo, Via Cintia - I-80126 Napoli,
Italy}

\date{\today}

\begin{abstract}
Recent studies have provided new experimental information on the very neutron-rich nucleus $^{135}$Sb.
We have performed a shell-model calculation for this nucleus using a realistic effective interaction derived from 
the CD-Bonn nucleon-nucleon potential. This gives a very good description of the observed properties of $^{135}$Sb.
We show that the anomalously low position of the first excited state, $J^{\pi}=5/2^{+}$, and the strongly hindered $M1$
transition to the ground state have their origin in the effective neutron-proton interaction.
\end{abstract}    

\pacs{21.60.Cs, 21.30.Fe, 27.60.+j}

\maketitle

The nucleus $^{135}$Sb has been investigated  through the $\beta$ decay of $^{135}$Sn in two recent 
studies \cite{korg01,sher02}.  The former, performed at OSIRIS/STUDSVIK, has  
produced  $^{135}$Sn via fast neutron fission of a $^{238}$U target inside a special 
ion source, 
while the latter, at ISOLDE/CERN, has made use of $^{135-137}$Sn nuclei isolated by selective
laser ionization.   
Motivated by astrophysical interest, these studies aimed 
to acquire new data on the decay properties of Sn isotopes 
beyond doubly magic $^{132}$Sn,  which are of great interest for the $r$ process modeling.
It is not of  minor interest, however,  to 
access  exotic nuclei, as Sb isotopes with $N>82$, in order to explore for possible 
changes in nuclear structure properties when moving towards the neutron drip line.
Actually, in \cite{korg01,sher02} the level structure of $^{135}$Sb has been studied. This nucleus, 
with a  $N/Z$ ratio of 1.65, is at present the most exotic nucleus beyond $^{132}$Sn for which information exists
on excited states.

A level scheme of $^{135}$Sb, whose ground state has   $J^{\pi}=7/2^{+}$,  
was firstly proposed in \cite{bhat98}, where, using prompt 
fission from $^{248}$Cm, four excited states 
at  an energy of 707, 1118, 1343, and 1973 keV were identified  and given spin-parity  
assignment $J^{\pi}= 11/2^{+}$,  $15/2^{+}$, $19/2^{+}$, and  $23/2^{+}$, respectively.
A shell-model calculation \cite{bhat98} with  an empirical two-body interaction 
interpreted the first three 
states as originating from
the $\pi g_{7/2} (\nu {f_{7/2}})^{2}$ configuration while the $23/2^{+}$ state was 
attributed to the $\pi g_{7/2} \nu f_{7/2} h_{9/2}$ configuration. 

In \cite{sher02} several new excited states have been identified, 
three  of them having also been observed in \cite{korg01}. In particular, both these studies have found 
a first  excited state at 282 keV and assigned to it spin and parity $5/2^{+}$.
None of the other states  has received in \cite{korg01,sher02} spin-parity assignment.
In the very recent work of Ref. \cite{sher05a}, tentative spin and parity attribution has been proposed for 
some of these states and the position of new levels has been established, including the
$J^{\pi}=3/2^{+}$ and  $9/2^{+}$ yrast states at 440 and 798 keV, respectively. 

Looking at the systematics of the lowest-lying $5/2^{+}$ state in   
odd Sb isotopes, it 
appears that this state falls down in energy in $^{135}$Sb. 
The authors of Ref. \cite{sher02}  suggest   that the  $5/2^{+}$ state
retains more single-particle character than the  $7/2^{+}$ 
ground state,  based on their estimates of $\log ft$ values. 
In this context, the low position of the $5/2^{+}$ state is quite unexpected and is viewed
as a downshift of the $d_{5/2}$ proton level  relative to the $g_{7/2}$ one, as a consequence of  
a more diffuse nuclear  surface produced by 
the two neutrons beyond the 82 shell closure. 
In \cite{sher02} results of a shell model calculation for $^{135}$Sb are also presented.
We shall comment on them later.

In the very recent work of Refs. \cite{mach05,korbe} the nature of the $5/2^{+}$ state in $^{135}$Sb
has been further investigated. In particular, at OSIRIS/STUDSVIK the Advanced 
Time-Delayed $\beta \gamma \gamma$($t$) method has been used to measure the lifetime of this state.
A very small upper limit for the $B(M1)$ was found, thus 
evidencing a strongly hindered transition.
This was seen as  a confirmation of the single-particle nature of the
$5/2^{+}$ state.
 
In this paper we report on  a shell-model study of $^{135}$Sb employing  matrix elements of the
two-body effective interaction derived from a modern nucleon-nucleon ($NN$) potential. It is our main aim to
verify  to what extent a realistic shell-model calculation
can account for the properties of $^{135}$Sb,  with special attention to the $5/2^{+}$ state, 
and try understand if there is a real need of shell structure modifications to explain the experimental data.

We assume that $^{132}$Sn is a closed core and let the valence
neutrons occupy the six levels $0h_{9/2}$, $1f_{7/2}$, $1f_{5/2}$, $2p_{3/2}$,
$2p_{1/2}$, and  $0i_{13/2}$ of the 82-126 shell, while for the proton
the model space includes the five  levels  $0g_{7/2}$, $1d_{5/2}$, $1d_{3/2}$, $2s_{1/2}$,
and $0h_{11/2}$ of the 50-82 shell. 

The two-body matrix elements of the effective interaction are derived from the CD-Bonn $NN$ 
potential \cite{mac01}. The strong short-range repulsion of the latter is renormalized by means of the 
new approach of Ref. \cite{bog02}, which has proved to be an advantageous
alternative to the usual $G$-matrix method \cite{bog02,gm}. In this approach, a smooth
potential, $V_{\rm low-k}$, is constructed by integrating out the high-momentum components,
i.e. above a  certain cut-off momentum $\Lambda$,  of the
bare $NN$ potential $V_{NN}$. The $V_{\rm low-k}$ preserves the physics
of $V_{NN}$ up to $\Lambda$
and  can be used directly in the calculation of shell-model effective interactions.
In the present paper, we have used  for  $\Lambda$ the value 2.2 fm$^{-1}$.

Once the $V_{\rm low-k}$ is obtained, the calculation of the effective
interaction is carried out within the framework of  the $\hat {Q}$-box plus folded diagram method.
A description of this method 
can be found in \cite{ei}. Here, we only mention that 
the $\hat{Q}$ box is calculated including  diagrams up
to second order in $V_{\rm low-k}$. The computation  of these diagrams is 
performed within the harmonic-oscillator basis using  intermediate states composed of all possible hole 
states and particle states restricted  to the five shells above the Fermi surface.
The oscillator parameter used is $\hbar \omega = 7.88$ MeV and for protons the Coulomb force has been
explicitly added to the $V_{\rm low-k}$ potential. 

As regards the single-particle (SP) energies,
they have been taken from experiment. In particular, the spectra  \cite{nndc} 
of $^{133}$Sb and $^{133}$Sn have been used to fix the proton
and neutron SP energies, respectively. 
The only exceptions are the the proton $\epsilon_{s_{1/2}}$ and neutron $\epsilon_{i_{13/2}}$,
whose corresponding levels are still missing. Their values have been taken from Refs. \cite{and97} and \cite{cor02},
respectively, where it is discussed how they are determined.
For the sake of completeness, the adopted SP energies relative to $^{132}$Sn  are reported in Table I.

\begin{table}[H]
\caption{Proton and neutron single-particle energies (in MeV).} 

\begin{ruledtabular}
\begin{tabular}{cccc}
$\pi(n,l,j)$&$\epsilon$ & $\nu (n,l,j)$& $\epsilon$\\
\colrule
$0g_{7/2}$ & -9.66 & $1f_{7/2}$ & -2.45\\
$1d_{5/2}$ & -8.70 & $2p_{3/2}$ & -1.60 \\
$2d_{3/2}$ & -7.22 & $0h_{9/2}$ & -0.89\\
$0h_{11/2}$ & -6.87  & $2p_{1/2}$ & -0.80\\
$2s_{1/2}$ &  -6.86 & $1f_{5/2}$ & -0.45\\
&   &                $0i_{13/2}$ & 0.24  \\
\end{tabular}
\end{ruledtabular}
\end{table}

The needed mass excesses are taken from \cite{foge99}.

We now present the results of our calculations, which have been carried out
by using the OXBASH shell-model code \cite{oxba}. Let us start with  the binding energy of
the ground state. Our calculated value is $16.411 \pm0.074$ MeV, which compares very well 
with  the experimental one,  
$16.585 \pm 0.104$ MeV \cite{audi03}.  
Note that the error on the calculated value arises from the experimental errors on  
the proton and neutron separation energies of $^{133}$Sb and $^{133}$Sn \cite{foge99}.

The experimental \cite{sher05a,nndc} and
calculated spectrum  of  $^{135}$Sb are compared in Fig. 1, where we have reported  the  
experimental yrast levels which have received spin-parity assignment  and the calculated yrast states
with  the same spin and parity. We see that the agreement between theory and experiment is very 
good, the largest discrepancy, about 300 keV,  occurring for the $23/2^{+}$ state. It is worth noting that our
calculation overestimates the observed $5/2^{+}$ state by only 100 keV. This result becomes more relevant
if we consider the outcome  of  the previous shell-model calculations  of Refs. \cite{sher02} and
\cite{sark04}. 
In these two calculations different two-body matrix elements have been used:
in \cite{sark04} an empirical interaction was obtained by modifying the CW5082 interaction of 
Chou and Warburton \cite{chou92}, while in Ref. \cite{sher02}  the  effective 
interaction was derived 
from the CD-Bonn $NN$ potential by means of a  $G$-matrix folded-diagram method including diagrams up 
to third order.   Both calculations, however, 
predict the $5/2^{+}$ state at  about 300 keV  
above the experimental one.  To overcome this difficulty, in \cite{sher02} a down-shift of 
the proton $d_{5/2}$ energy was proposed. 

\begin{figure}[H]
\begin{center}
\includegraphics[scale=0.9,angle=0]{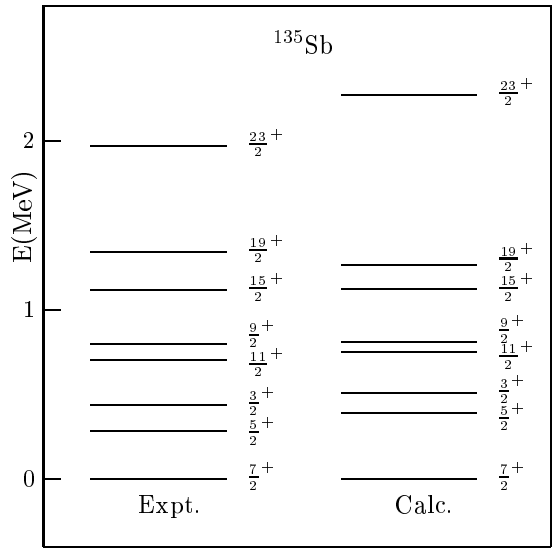}
\end{center}
\caption{Experimental and calculated spectrum of $^{135}$Sb.}
\end{figure}

It is worth mentioning that in the preliminary calculations   \cite{mach05,korg05} we too could not 
account for the low-lying  $5/2^+$ state in $^{135}$Sb, our result coming close to that of \cite{sher02}.
Actually, the effective interaction used in the present calculation differs from that of 
\cite{mach05,korg05} in that a larger number of intermediate states is used in its derivation.
More precisely, in  \cite{mach05,korg05} the computation of the $\hat Q$-box diagrams was
performed including intermediate states composed of particle and hole states restricted to two major shells 
above and
below the Fermi surface. Furthermore, the Coulomb 
interaction was not considered and a smaller  value 
of the cutoff momentum ($\Lambda =2.1$ fm$^{-1}$)  was used.  In a future publication we shall
discuss the connection between the dimension of the intermediate-state space and  
the optimal value of $\Lambda$.  
We have verified  that while the present effective interaction is generally not 
very different from the old one, significant changes occur in some  neutron-proton matrix elements.   
In particular, this is the case  of those of the $\pi f_{7/2} \nu g_{7/2}$ and   $\pi d_{5/2} \nu g_{7/2}$ 
configurations,
which, as we shall see later, play a crucial role in the structure of the $5/2^{+}$ state of $^{135}$Sb. 
In this connection, it is worth noting that a direct test of these matrix elements is provided by $^{134}$Sb.  
We have found that that our present  effective interaction yields energies of all the
members of the  $\pi f_{7/2} \nu g_{7/2}$ configuration which are in very good agreement 
with the eight  lowest observed states \cite{sher05}.  
Note that with our previous interaction the agreement was not of the same quality \cite{garg04}.
As for the   $\pi d_{5/2} \nu g_{7/2}$ multiplet, only two members   
have been identified, the $1^{-}$ and  $2^{-}$ states at 0.885 and 0.935 MeV excitation energy, respectively. Both of them
are very well reproduced by our calculation. 

Let us now come back to $^{135}$Sb. Our  wave functions for the ground state and 
the first excited $5/2^{+}$ state are

\begin{equation}
|\frac{7}{2}^{+} \rangle = 0.87 | \pi g_{7/2} (\nu f_{7/2})^{2} \rangle + \cdots ,\\
\end{equation}

\begin{equation}
|\frac{5}{2}^{+} \rangle =  0.67 | \pi d_{5/2} (\nu f_{7/2})^{2} \rangle + 0.48 | \pi g_{7/2} (\nu f_{7/2})^{2} \rangle +\cdots,
\end{equation}
where we have omitted components having a weight less than $5\%$. We see that the $5/2^+$ state 
has a large weight of the $\pi d_{5/2} (\nu f_{7/2})^{2}$ 
configuration. It is important 
to point out, however, that this does not arise, as 
it would be expected, from the pairing correlation between the two like nucleons. 
Rather, it is a consequence of the neutron-proton
interaction.
In fact, we have verified that switching off this  interaction  the 
$5/2^{+}$ state is dominated by the $\pi g_{7/2} (\nu f_{7/2})^{2}$ configuration. This is because the neutron-neutron 
interaction in  the $ (\nu f_{7/2})^{2}_{J^{\pi}=2^{+}}$ state  is very  attractive, 
the difference with the 
$ (\nu f_{7/2})^{2}_{J^{\pi}=0^{+}}$ matrix element being  smaller than the 
$\epsilon_{d_{5/2}}-\epsilon_{g_{7/2}}$ spacing.    
This is confirmed by the low excitation energy of the $2^{+}$ state observed in $^{134}$Sn. The neutron-proton
interaction largely increases the weight of the $\pi d_{5/2} (\nu f_{7/2})^{2}$ configuration 
in the $5/2^{+}$ state
since it is on the average more attractive in the 
$\pi d_{5/2} (\nu f_{7/2})^{2}$ than in the $\pi g_{7/2} (\nu f_{7/2})^{2}$ configuration.
Of course, for the latter no contribution comes from the states  
$(\pi g_{7/2} \nu f_{7/2})_{J^{\pi}=0^{+},7^{+}}$, as they cannot couple to the $f_{7/2}$ neutron 
to give $J^{\pi}=5/2^{+}$. 
We show in Table II the diagonal matrix 
elements of the effective interaction for these two configurations. 

\begin{table}[H]
\caption{Diagonal matrix elements of the neutron-proton effective interaction (in MeV) for
the $\pi f_{7/2} \nu g_{7/2}$ and $\pi d_{5/2} \nu g_{7/2}$ configurations.}
\begin{ruledtabular}
\begin{tabular}{ccc}
$J$&  $\pi f_{7/2} \nu g_{7/2}$   &  $\pi d_{5/2} \nu g_{7/2}$ \\
\colrule
0 & -0.58 &  \\
1 & -0.59 & -0.49 \\
2 & -0.21 & -0.26 \\
3 & -0.22 & -0.11 \\
4 & -0.04 & -0.19 \\
5 & -0.16 & -0.04  \\
6 & 0.04  & -0.59  \\
7 & -0.31  &    \\
\end{tabular}
\end{ruledtabular}
\end{table}

It may be also of interest to see how the effective interaction affects the centroids of the $g_{7/2}$
and $d_{5/2}$ proton strengths.  Our calculated values are -10.06  and -9.35 MeV for the former and the latter, respectively, both
of them relative to the calculated ground-state energy of $^{134}$Sn. In this way, the displacements 
from the input values -9.66 and -8.70 MeV (see Table I) can be  directly related to the proton-neutron effective interaction.
We find, therefore, that  the spacing between the two centroids is 0.71 MeV to be compared to the proton SP 
splitting, 0.96 MeV. 
Note that in \cite{sher02} the effective interaction produced an increase in  such spacing of about 100 keV.
As already mentioned, a decrease of the $d_{5/2} - g_{7/2}$ splitting of 0.300 MeV was therefore called for in \cite{sher02}. 

We now discuss our prediction for the $5/2^{+} \rightarrow 7/2^{+}$ $M1$ transition.  As shown in (2), 
our wave function
for the $5/2^{+}$ state contains a sizeable  component of the 
$\pi g_{7/2} (\nu f_{7/2})^{2}$ configuration, which on the other hand dominates  the $7/2^{+}$ ground state. This makes the
theoretical  
$B(M1; 5/2^{+} \rightarrow 7/2^{+})$ larger than  the experimental value  when the free $M1$ operator is used.
From the measured half-life of the $5/2^{+}$ level, $T_{1/2}=6.0(7)$ ns, 
an upper limit of $0.29\cdot10^{-3} \, \mu_{N}^{2}$ was deduced for the $B(M1)$ transition rate \cite{mach05,korg05}.
Our calculated value,  
$25 \cdot  10^{-3} \, \mu_{N}^{2}$, is about 100 times larger. However, as is well known, the magnetic 
operator in the nucleus may be significantly different from the free
operator owing to core-polarization effects and mesonic exchange currents. Here, we have calculated
the matrix elements of effective $M1$ operator taking into account first order diagrams in $V_{\rm low-k}$. 
Using these  matrix elements, 
the transition rate for  the $5/2^{+}$ state turns out to be  $4 \cdot  10^{-3} \, \mu_{N}^{2}$, which 
is only an order of magnitude larger than 
the experimental value.
The reason for this decrease is that we now have a non-zero off-diagonal matrix element between   
the  $d_{5/2}$ and $g_{7/2}$ proton levels which is opposite in sign to the
diagonal $g_{7/2}$ matrix element. In other words, both components of the $5/2^{+}$ state contribute to the  
$B(M1)$ value and their contributions partially compensate each other.
The balance between these two components is a delicate one, being very sensitive to  
small changes in the wave functions of the involved states or in the effective $M1$ operator. 
In this connection, it should be pointed out that in our calculation we do not consider any meson-exchange correction.
We may therefore consider that  
the agreement between the experimental and calculated $B(M1)$ values is quite satisfactory. It is important to note 
that  
the experimental magnetic moment of the $7/2^{+}$ ground state is $3.0 \, \mu_{N}$, to be compared to the calculated
value $1.7 \, \mu_{N}$ when  the free operator is used and to $2.5 \, \mu_{N}$ with the effective $M1$ operator.

In summary, our realistic shell-model calculation gives a very good description of the observed
spectroscopic properties of  exotic $^{135}$Sb. This is obtained by employing an effective interaction derived from the CD-Bonn
$NN$ potential within a folded-diagram method including diagrams up to second order in $V_{\rm low-k}$.
A crucial  role is played by the proton-neutron interaction, whose effects  have been tested on the proton-neutron nucleus $^{134}$Sb. 
In particular, we have found that this interaction makes the $5/2^{+}$ state in $^{135}$Sb of admixed nature, which can
explain its low excitation energy as well the highly hindered 
$5/2^{+} \rightarrow 7/2^{+}$ $M1$ transition.
We have shown that in  our shell-model calculation there is no need to introduce an {\it ad-hoc} decrease of the
proton $d_{5/2}$ SP energy, the effective interaction
properly accounting for the observed effects.

\begin{acknowledgments}
This work was supported in part by the Italian Ministero
dell'Istruzione, dell'Universit\`a e della Ricerca  (MIUR).
\end{acknowledgments}


\begin{thebibliography}{9}
\bibitem{korg01} A. Korgul, H. Mach, B. Fogelberg, W. Urban, W. Kurcewicz, and V. I. Isakov, Phys. Rev. C {\bf 64},
021302(R) (2001).
\bibitem{sher02} J. Shergur {\it et al.}, Phys. Rev. C {\bf 65},
034313 (2002).
\bibitem{bhat98} P. Bhattacharyya {\it et al.}, Eur. Phys. J. A {\bf 3},
109 (1998).
\bibitem{sher05a} J. Shergur {\it et al.}, Phys. Rev. C {\bf 72},
024305 (2005).
\bibitem{mach05} H. Mach {\it et al.}, in {\it Key  Topics in Nuclear Structure},
Proceedings of the Eighth International Spring Seminar on Nuclear Physics, Paestum, 2004,
edited by A. Covello (World Scientific, Singapore, 2005), p. 205.
\bibitem{korbe} A. Korgul {\it et al.} (to be published).
\bibitem{mac01} R. Machleidt, Phys. Rev. C {\bf 63}, 024001 (2001).
\bibitem{bog02} S. Bogner, T. T. S. Kuo, L. Coraggio, A. Covello, and N. Itaco, Phys. Rev.
C {\bf 65}, 051301 (2002).
\bibitem{gm} A. Covello, L. Coraggio, A. Gargano, N. Itaco, and T. T. S. Kuo, in {\it Challenges of Nuclear Structure},
Proceedings of the Seventh International Spring Seminar on Nuclear Physics, Maiori, 2001,
edited by A. Covello (World Scientific, Singapore, 2002), p. 139.
\bibitem{ei} T. T. S. Kuo and E. Osnes, {\it Lecture Notes in Physics}, Vol. 
364, (Springer-Verlag, Berlin, 1990).
\bibitem{nndc} Data extracted using the NNDC On-line Data Service from the ENSDF database,
file revised as of July 11, 2005.
\bibitem{and97} F. Andreozzi, L. Coraggio, A. Covello, A. Gargano, T. T. S. Kuo, 
and A. Porrino,  Phys. Rev. C {\bf 56}, R16 (1997).
\bibitem{cor02} L. Coraggio, A. Covello, A. Gargano, and N. Itaco, Phys. Rev.
C {\bf 65}, 051306(R) (2002).
\bibitem{foge99} B. Fogelberg, K. A. Mezilev, H. Mach, V. I. Isakov, and J. Slivova
       Phys. Rev. Lett. {\bf 82}, 1823 (1999).
\bibitem{oxba}B. A. Brown, A. Etchegoyen, and W. D. M. Rae, The computer code OXBAH, MSU-NSCL, Report No. 534.
\bibitem{audi03} G. Audi, A. H. Wapstra, and C. Thibault, Nucl. Phys. {\bf A729}, 337 (2003).
\bibitem{sark04} S. Sarkar and M. S. Sarkar, Eur. Phys. J. A {\bf 21}, 61 (2004).
\bibitem{chou92} W. T. Chou and E. K. Warburton, Phys. Rev. C {\bf 45}, 1720 (1992).
\bibitem{korg05} A. Korgul {\it et al.}, Eur. Phys. J. A direct (2005).
\bibitem{sher05} J. Shergur {\it et al.}, Phys. Rev. C {\bf 71},
064321 (2005).
\bibitem{garg04} A.  Gargano, Eur. Phys. J. A {\bf 20}, 103 (2004).
\end{thebibliography}
\end{document}